\documentclass{iau}
\usepackage{graphicx}

\title[] 
{Birth accelerations of neutron stars}

\author[R. Heras]   
{Ricardo Heras}

\affiliation{Preparatoria Abierta, SEIEM en Toluca Edo. de Mexico\\email: {\tt ricardoherasosornor@gmail.com}}

\pubyear{2012}
\volume{291}  
\jname{\mbox{Neutron Stars and Pulsars: Challenges and Opportunities after 80 years}}
\editors{J. van Leeuwen, ed.} 
\begin{document}

\maketitle
\begin{abstract}
We suggest that neutron stars experienced at birth three related
physical changes, which may  originate in magneto-rotational
instabilities: (i) an increase
in period from the initial value $P_0$ to the current value $P_s$, implying a change of rotational energy $\Delta E_{\rm rot}$; (ii) an exponential decay of its magnetic field from the initial value $B_0$ to the current surface value $B_s$, implying a change of radiative energy $\Delta E_{\rm rad}$; and (iii) an increase of space velocity from the initial value $v_0$ to the current value $v$,  implying a change of kinetic energy $\Delta E_{\rm kin}$. These changes are assumed to be connected by $\Delta E_{\rm rad}+ \Delta E_{\rm kin}\!=\! \Delta E_{\rm rot}.$ This means that the radiation loss and increase of kinetic energy are both at the expense of a rotational energy loss. It is shown that this energy conversion occurs during times of order of $10^{-4}$\,s if the neutron stars are born with magnetic fields in the range of $10^{15}\!\!-\!\!10^{16}$\,G and initial periods in range $1\!-\!20$\,ms. It is shown that the birth accelerations of neutron stars are of the order of $10^{8}$g.
\end{abstract}

\firstsection 
\section{Introduction}
The idea that neutron stars are born with magnetic fields typical of magnetars $(10^{15}\!-\!10^{16}$G) and periods typical of millisecond pulsars $(1\!-\!20$ ms) is based on the assumption that neutron stars experienced three related physical processes occurring at the end of their birth. Because of magneto-rotational instabilities occurring in neutron stars during their birth, these stars could experience an abrupt change of rotational energy which could cause a loss of radiative energy and a gain in kinetic energy:
\begin{equation}
\Delta E_{\rm rad}+ \Delta E_{\rm kin}\!=\! \Delta E_{\rm rot}.
\end{equation}
A similar energy conversion but with a different radiative term is the
basis of the ``Rocket Model" proposed by \cite[Harrison \& Tademaru
  (1975)]{Harrisontademaru01}. The idea of a loss of rotational energy
during the birth process of a neutron star was already considered by
\cite[Usov (1992)]{usov01} for the case of millisecond pulsars. On the
other hand, \cite[Spruit (2008)]{spruit01} has suggested that a
differential rotation in the final stages of the core collapse process
can produce magnetic fields typical of magnetars. He has pointed out
that some form of magneto-rotational instability may be the cause of
an exponential growth of the magnetic field; but that once
formed in the core collapse, this magnetic field may decay
again through magnetic instabilities. In connection with the idea that
neutron stars are born as magnetars, \cite[Geppert \& Rheinhardt
  (2006)]{geppertrheinhart01} have discussed a magneto-hydrodynamical
process that significantly reduces the initial magnetic field of a
newly-born neutron star in fractions of a second. According to these
authors, such a field reduction is due to magneto-hydrodynamical
instabilities, which seem to be inevitable if neutron stars are born
as magnetars.

The present work focusses on the initial dynamics of neutron stars. Specifically, by considering Eq.~(1.1) and the assumption that neutron stars are born with magnetic fields of magnetars and periods of millisecond pulsars, the birth acceleration of neutron stars will be estimated.

\section{The birth-ultrafast-magnetic-field-decay model of neutron stars}

The Larmor formula for the power radiated by a time-varying magnetic
dipole moment  $P=2\ddot{\mu}^2/(3c^3)$,  and the estimate
$\ddot{\mu}\!\sim\!\mu_0/\tau^2$, where $\tau$ is the characteristic time in the exponential field decay law $B(t)\!=\!B_0e^{-t/\tau},$ imply the equation
$P\!\simeq \!2\mu_0^2/(3c^3\tau^4),$ which can be used together  with the relation $\mu_0\!=\!B_0R^3/2$ to yield the power radiated by a neutron star of radius $R$ and an initial magnetic field $B_0$:
\begin{equation}
P\!\simeq \!\frac{B_0^2 R^6}{6c^3\tau^4}.
\end{equation}
Consider now the specific time $\tau_s$ elapsed during the field decay from the initial value $B_0$ to the current surface magnetic field $B_s$. The condition $B(\tau_s)\!=\!B_s$ and the law $B(t)\!=\!B_0e^{-t/\tau}$ imply $B_s\!=\!B_0e^{-\tau_s/\tau}$ or equivalently
\begin{equation}
\tau_s\!=\!\tau\ln(B_0/B_s).
\end{equation}
From Eqs.~(2.1) and (2.2) it follows that the change of electromagnetic energy radiated $\Delta E_{\rm rad}\!\simeq \tau_s P$ is explicitly given by
\begin{equation}
\Delta E_{\rm rad}\!\simeq \frac{\!B_0^2 R^6\ln(B_0/B_s)}{6c^3\tau^3}.
\end{equation}
The initial magnetic field $B_0$ is associated with the initial
angular frequency $\Omega_0$ (or equivalently with its associated
initial period $P_0$). The initial rotational kinetic energy of a
neutron star is $I\Omega_0^2/2$, where $I=2MR^2/5$ is the moment of inertia
for a neutron star with mass $M$. The rotational kinetic energy associated with the surface magnetic field $B_s$ is  $I\Omega_s^2/2$, where $\Omega_s$ is the current angular frequency (and $P_s$ its associated period) of the neutron star. It is expected that the angular frequency of neutron stars decreases during the field decay of this star. This means that $\Omega_0>\Omega_s$ and so the rotational kinetic energy must decrease. Since $\Omega=2\pi/P$ it follows that the period increases during the birth process ($P_s>P_0$).
By assuming that the change of radiative energy in Eq.~(2.3) and the change of kinetic energy $E_{\rm kin}=Mv^2/2$ of the neutron star (the initial velocity is taken to be zero)
are both at the expense of the change of rotational energy $ \Delta E_{\rm rot}\!=\!4\pi^2MR^2(P_0^{-2}\!\!-\!\! P_s^{-2})/5$ the energy conservation reads
\begin{equation}
\underbrace{\frac{B_0^2R^6\ln(B_0/B_s)}{6c^3\tau^3}}_{\Delta E_{\rm rad}}\!+\! \underbrace{\frac{M v^2}{2}}_{\Delta E_{\rm kin}}\!=\! \underbrace{\frac{4\pi^2MR^2}{5}\bigg(\frac{1}{P_0^2}\!-\!\frac{1}{P_s^2}\bigg)}_{\Delta E_{\rm rot}}.
\end{equation}
To get an idea of the order of the time $\tau$, consider the Crab pulsar B0531+21 which has $M\!= \!1.4 M_\odot$; $R\!=\!10$ km; $P_s\!=\!0.033$ s$; B_s\!=\!3.78\!\times\!10^{12}$G; $ v_{\perp}\!=\!141$ km/s which implies
$v\!=\!172.7$ km/s, where the relation $v\!\approx\!\sqrt{3/2}\:v_{\perp}$ has been used (\cite[Lyne \& Lorimer 2006]{Lyne01}). It has been suggested that the Crab pulsar was born with $P_0\!\approx\!0.019$ s (\cite[Lyne \etal\ 1993]{Lyne02}). Using this initial period and assuming $B_0\!=\!5.8\times\!10^{15}$\,G, Eq.~(2.4) yields $\tau\! \approx R/c.$ For the magnetar J1809-1943, Eq.~(2.4) also implies $\tau\! \approx R/c$ if $P_0\!\approx\!0.02$ s and  $B_0\!=\!9.5\times\!10^{15}$G. For the isolated millisecond pulsar MSP B1257+12, Eq.~(2.4) yields $\tau\! \approx R/c$ if  $P_0\!\approx\!0.059$ s and  $B_0\!=\!2.7\times\!10^{15}$G. The time $\tau=R/c$ agrees with the idea that neutron stars are born with magnetic fields in the range of $10^{15}\!\!-\!\!10^{16}$G and initial periods in range of $1\!-\!20$ ms. Magnetic field decays from one to eight orders of magnitude satisfy $2.3  \leq\ln(B_0/B_s)\leq4.4$. When this relation and $\tau=R/c$ are used in Eq.~(2.2) one has
\begin{equation}
\tau_s\!\sim\!10^{-4}\; {\rm s}.
\end{equation}
This means that the field decay from $B_0$ to $B_s$ is ultrafast if $B_0$ lies in the range of $10^{15}\!\!-\!\!10^{16}$G and $P_0$ in the range of $1-20$ ms.

\section{Birth accelerations of neutron stars}
The energy conversion in Eq.~(2.4) occurs suddenly and therefore one can write
\begin{equation}
v\approx a\tau_s,
\end{equation}
where $a$ is the birth acceleration. Equations~(2.2), (2.4), (3.1) and $\tau=R/c$ yield
\begin{equation}
a= \sqrt{\frac{8\pi^2c^2(P_0^{-2}- P_s^{-2})}{5\ln{(B_0/B_s)^2}}-\frac{B_0^2 R c^2}{3M\ln{(B_0/B_s)}}}.
\end{equation}
Using this equation with $M\!= \!1.4 M_\odot$ and $R\!=\!10$ km, and taking the specific values for $B_s$ and $P_0$ from the ATNF Pulsar Catalogue and assuming values for $P_0$ and $B_0$, the birth acceleration can be calculated and expressed in terms of g=9.8 m/s$^{2}$. Four sets of neutron stars with reported transverse velocities will be now considered:

\vspace{1ex}
\noindent {Interval: $2$ s $\!<\!P_s\!\leq\!8.5$ s.}\\
There are 9 neutron stars in this interval. The average values are ${P_s}=4.82$ s and ${B_s}=3.14\times 10^{13}$G. If the values $P_0=0.02$ s and $B_0\!=\! 7\times \!10^{15}$G are assumed then Eq.~(3.2) predicts the birth acceleration $a=5.0\times10^{8}$g.

\vspace{1ex}
\noindent {Interval: $1$ s $\leq P_s\leq 2$ s.}\\ There are 29 neutron stars in this interval. The average values are ${P_s}=1.32$ s, ${B_s}=4.17\times 10^{12}$ G. If the values $P_0=0.02$ s and $B_0=5\times 10^{15}$G are assumed then Eq.~(3.2) gives the birth acceleration $a=5.8\times10^{8}$g.

\vspace{1ex}
\noindent {Interval: $0.02$ s $ \leq P_s< 1$ s.}\\
 There are 130 neutron stars in this interval. The average values are ${P_s}=0.41$ s and  ${B_s}=1.28\times 10^{12}$G. If the values ${P_0}\!=\!0.02$ s and $B_0=5\times 10^{15}$G are assumed then Eq.~(3.2) implies the birth acceleration $a=4.5\times10^{8}$g.

\vspace{1ex}
\noindent {Interval: $0.0015$ s $\! \leq P_s<\! 0.009$ s.}\\
 There are 9 millisecond pulsars in this interval. The average values are ${P_s}\!=\!0.005$ s and ${B_s}=2.59\times 10^{8}$G. If the values ${P_0}\!=\!0.0049$ s and $B_0\!=\!10^{15}$G then Eq.~(3.2) predicts the birth acceleration
$a=3.1\times10^{8}$g.

\vspace{1ex}
It can be concluded that birth accelerations of neutron stars occur in times $\tau_s\!\sim\!10^{-4}$\,s and are of the order of $10^{8}$\,g.
If for example, $a=5\times 10^{8}$\,g and $\tau_s\!\sim\!10^{-4}$\,s are assumed then the birth velocity (kick velocity) is of the order of 500 km/s. This value is approximately the average value for pulsar velocities found by \cite[Lyne \& Lorimer  (2006)]{Lyne01}.

\end{document}